\documentclass[runningheads]{llncs}

\usepackage[T1]{fontenc}
\usepackage{graphicx}
\usepackage{subfigure}
\usepackage{amsmath}
\usepackage{multirow}
\usepackage{booktabs} 
\usepackage{array} 
\usepackage{enumitem}
\usepackage{hyperref}
\usepackage{marvosym}
\usepackage{lipsum}

\usepackage{amsmath,amssymb}
\usepackage{booktabs}
\usepackage{subfigure}
\usepackage{caption}
\usepackage{xcolor}
\usepackage{subcaption}
\usepackage{float}

\begin{document}

\title{Diff3Dformer: Leveraging Slice Sequence Diffusion for Enhanced 3D CT Classification with Transformer Networks}

\author{
Zihao Jin\inst{1}$^{*}$\and  
Yingying Fang\inst{2}$^{\dag*}$\and  
Jiahao Huang\inst{3} \and 
Caiwen Xu\inst{3} \and 
Simon Walsh\inst{2} \and 
Guang Yang\inst{2,3,4,5}$^{\dag}$} 
\authorrunning{Z. Jin et al.}
\titlerunning{Diff3Dformer}

\renewcommand{\thefootnote}{}
\footnotetext{$^*$Z. Jin and Y. Fang---Equal contribution.}

\institute{
Department of Metabolism, Digestion and Reproduction, Imperial College London, London, UK \and
National Heart and Lung Institute, Imperial College London, London, UK  \and
Bioengineering Department and Imperial-X, Imperial College London, London, UK  \and
Cardiovascular Research Centre, Royal Brompton Hospital, London, UK
\and
School of Biomedical Engineering \& Imaging Sciences, King's College London, London, UK
\\
\email{\{y.fang;g.yang\}@imperial.ac.uk} 
\\
}

\maketitle            

\begin{abstract}
The manifestation of symptoms associated with lung diseases can vary in different depths for individual patients, highlighting the significance of 3D information in CT scans for medical image classification. While Vision Transformer has shown superior performance over convolutional neural networks in image classification tasks, their effectiveness is often demonstrated on sufficiently large 2D datasets and they easily encounter overfitting issues on small medical image datasets. To address this limitation, we propose a Diffusion-based 3D Vision Transformer (Diff3Dformer), which utilizes the latent space of the Diffusion model to form the slice sequence for 3D analysis and incorporates clustering attention into ViT to aggregate repetitive information within 3D CT scans, thereby harnessing the power of the advanced transformer in 3D classification tasks on small datasets. Our method exhibits improved performance on two different scales of small datasets of 3D lung CT scans, surpassing the state of the art 3D methods and other transformer-based approaches that emerged during the COVID-19 pandemic, demonstrating its robust and superior performance across different scales of data. Experimental results underscore the superiority of our proposed method, indicating its potential for enhancing medical image classification tasks in real-world scenarios. 
The code will be publicly available at \url{https://github.com/ayanglab/Diff3Dformer}.

\end{abstract}

\section{Introduction}
3D volume analysis is crucial for the diagnosis or prognosis of lung diseases, as lesions can manifest at various depths in CT scans across different patients \cite{shamshad2023transformers}, such as those with COVID-19 or Interstitial Lung Disease. Compared to 2D analysis, 3D analysis enables a comprehensive examination of abnormal areas throughout the entire volume, offering a more thorough understanding of the patient's condition. Additionally, models analyzing 3D volumes eliminate the need for slice selection, leading to more efficient and reliable predictions compared to methods that rely on a limited number of preselected slices for patient-level decisions.

Given the imperative nature of these requirements, a quantity of 3D analysis methodologies emerged within the AI community during the COVID-19 pandemic for diagnosing and prognosticating patients \cite{fang2024post}. These methodologies can be broadly classified into aggregation (AG) methods \cite{zhang2021mia,mei2020artificial,miron2021covid}, 2.5D methods \cite{meng2023bilateral,hartmann2023covid}, and whole-scan methods (WS) \cite{he2020benchmarking,hou2021cmc,harmon2020artificial,wang2020weakly}. AG methods analyze 3D scans by aggregating results from all 2D slices \cite{zhang2021mia}, inherently limited in capturing intra-slice features. To overcome this limitation, WS methods input the entire scan into the model, allowing for comprehensive feature exploration throughout the 3D volume. Although 3D methods have demonstrated superior performance, they are susceptible to crashing due to overfitting, especially when dealing with a small dataset. As a compromise between diverse training samples and 3D features, 2.5D methods resample a fixed smaller number of slices from the entire scan, treating them as a unified input entity for network-based patient-level decision-making. While the resampling process enables extensive augmentation of the small dataset, the reliance on a randomly sampled subset of slices for patient-level decision-making still raises concerns among doctors who may potentially use these models in high-stakes contexts. Hence, enabling WS analysis within small datasets remains an urgent and unmet challenge.

Transformers have outperformed traditional CNN methods in vision classification tasks but require substantial data and memory resources, posing challenges for small datasets. 
Recent studies have illuminated the application of advanced Transformer architectures in 3D lung volume classification tasks within limited datasets \cite{hartmann2023covid,hsu2021visual,zhao2022prognostic,zhang2021mia}. 
To address memory constraints while handling high-dimensional 3D volumes, \cite{hartmann2023covid} employed 2.5D techniques, resampling 32 slices as input for the Timesformer model.  \cite{zhang2021mia} utilized the AG method, employing a 2D Swin Transformer to process CT volumes with varying slice counts. 
\cite{hsu2021visual,zhao2022prognostic} adopted a CNN-based preprocessing step to transform 3D volumes into sequences of low-dimensional CNN-based features, subsequently fed into Transformers for classification. 
To enhance performance on small datasets, \cite{zhao2022prognostic} employed Mixup  \cite{zhang2017mixup} data augmentation, while \cite{hartmann2023covid} and \cite{zhao2022prognostic}
explored transfer learning and self-supervised learning to achieve more general representations. Despite improvements brought by these methods, current Transformer-based 3D analysis still faces several limitations:  (1) the performance of   Transformer-based WS methods is still prone to overfitting and is significantly influenced by data scale; (2) there is a lack of comprehensive comparisons between Transformer-based and CNN-based models regarding their efficacy and robustness on small 3D datasets; and (3) interpreting the features used for patient-level decisions from these 3D scans remains underexplored in current research.

Motivated by prior work \cite{hsu2021visual,zhao2022prognostic}, our aim is to develop a robust 3D Transformer model that surpasses existing methods by effectively harnessing the global feature learning capabilities of transformers. Simultaneously, we strive to reduce data requirements and improve interpretability in 3D volume decisions.
To achieve this, we introduce the novel Diffusion-enhanced 3D Transformer (Diff3Dformer), which combines the advantageous latent space learning of the Diffusion Autoencoder with a Clustering Vision Transformer (ViT). This integration facilitates efficient feature extraction and information reduction during the global feature learning process.
The key contributions of this work are summarized as follows: 
(1). We discover how Clustering ViT can mitigate overfitting and effectively manage small datasets. 
(2) We introduce the Diffusion Autoencoder for self-supervised learning to extract semantically meaningful representations for enhanced 3D analysis. Additionally, we propose a novel pipeline that enables data-intensive Diffusion to be applied to small-scale 3D analysis using the efficient 3D solver Clustering ViT.
(3). We propose an interpretable slice fusion strategy to decode the model’s decisions into contributions from different clusters, enabling the explainability of the final patient-level decision from the Diff3Dformer.
(4). We conduct experiments on two different scales of small datasets, showcasing the robustness and consistent superiority of the proposed methods over different types of 3D analysis methods across varying medical image dataset scales. 

\section{Method}
\begin{figure}[t]
\centerline{\includegraphics[width=1\linewidth]{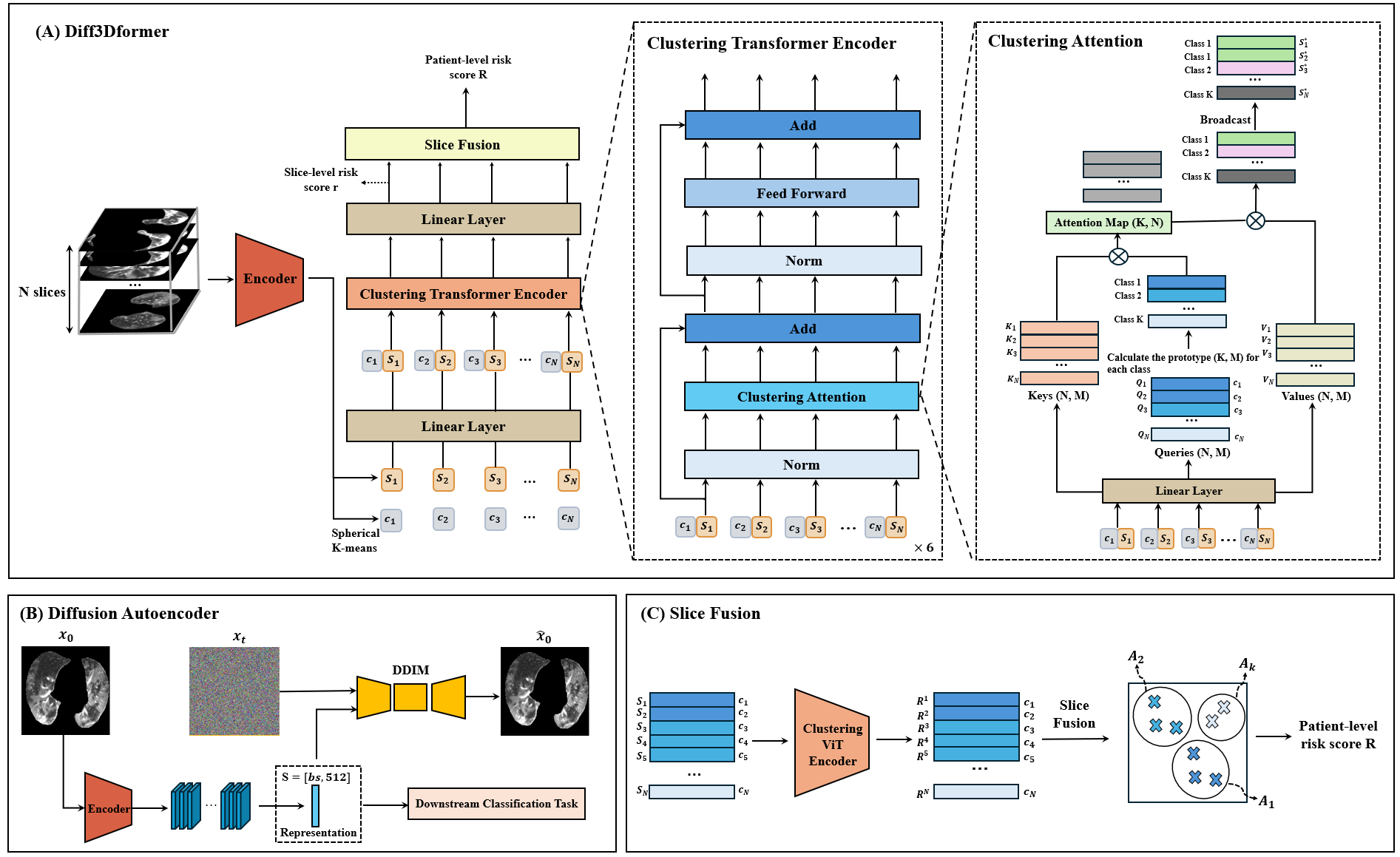}}
\caption{(A) The overview framework of Diff3Dformer. (B) The diffusion autoencoder is leveraged to learn a semantically meaningful representation by learning to reconstruct the 2D slice from a 512-dimensional representation and being used to represent CT volumes as a sequence of representations as the input of the clustering ViT model. (C) The slice fusion module provides final patient decisions and explanations of Diff3Dformer.
}
\label{Framework}
\end{figure}

The overview of Diff3Dformer is given in Fig.~\ref{Framework} (A). Prior to Diff3Dformer's prediction on individuals, we employ the encoder from the pretrained diffusion autoencoder to extract representations of each slice from CT volumes. By aggregating the slice representations from the entire dataset, we can learn slice prototypes (the centre of each cluster) specific to a particular disease using the spherical K-means method.
Given the learned prototype, Diff3Dformer starts by transforming a patient's 3D volumes into a sequence of the representations together with the cluster number it belongs to. The representation together with the assigned prototype number, they are fed into the Clustering ViT for global information learning through the self-attention map in the transformer. The cluster number here aids the model in detecting repetitive and similar patterns in the 3D volume, thereby reducing the number of features and enhancing computational efficiency within the traditional ViT.
Following the modification of slice representations, the final layer of DiffExplainer outputs scores for each slice to make the final decision. DiffExplainer employs global attention on predefined clusters, which are learned during training, to fuse slice scores from different clusters, thereby generating an explainable patient classification result.

\subsection{Representation Learning via Diffusion-based Autoencoder}

Motivated by recent strides in feature manipulation and disentanglement within Diffusion's latent space \cite{li2023your,preechakul2022diffusion,cho2023towards}, we are compelled to exploit these semantically meaningful features as representations for individual slices for downstream tasks. To derive highly meaningful representations for each slice, we utilized a Denoising Diffusion Implicit Model  (DDIM)-based autoencoder, as proposed by Preechakul et al. \cite{preechakul2022diffusion}, to reconstruct slices from CT scans. This autoencoder architecture consists of an encoder $\mathbf{E}$ and a DDIM model denoted as $\mathbf{D}$. To preserve meaningful information within the encoded representation from $\mathbf{E}$, DDIM is trained to reconstruct the original slice using this representation as a condition.

The models $\mathbf{E}$ and $\mathbf{D}$ are trained concurrently by optimizing the following loss function with respect to $\theta$ and $\phi$:
\begin{equation}
\min_{\phi, \theta} \mathcal{L}=\left\|\boldsymbol{\epsilon}-\mathbf{D}_\theta\left(\mathbf{x}_t, t, \mathbf{E}_\phi(\mathbf{x}_0)\right)\right\|_1,
\end{equation}
where  $\mathbf{x}_0$ represents any given slice and $\mathbf{x}_t$ is the  noise injected slice ($t$ iterations of Gaussian noise injection). The network $\mathbf{D}$ utilizes a UNet architecture consisting of layers of residual blocks, as described in \cite{dhariwal2021diffusion}. Meanwhile, the network $\mathbf{E}$ adopts the encoder architecture from $\mathbf{D}$.

Once the autoencoder achieves the optimal reconstruction quality, the encoder is separately utilized to extract the representations of each slice. These representations are then aggregated from each patient and clustered into $K$ clusters using Spherical K-means \cite{zhong2005efficient}. 
The clustering step will learn the potential prototypes of the slices within a specific dataset. These prototypes will further enable the quantification of the entire scan into a combination of prototype slices by grouping the slices with similar patterns together. Additionally, it will aid in reducing the features during self-attention learning in the subsequent Clustering ViT model introduced.

\subsection{Clustering ViT for 3D Classification}
After representation learning, each 3D volume can be transformed into a sequence of meaningful slice representations, each with its corresponding assigned cluster. Inspired by \cite{zhao2022prognostic,zheng2020end}, we introduce a clustering ViT model for 3D diagnostic and prognostic tasks based on the obtained slice sequence.

As illustrated in Fig. \ref{Framework} (A), the slice sequences obtained from each patient are padded to a fixed length $N$, mapped to $M$ dimensions using a linear layer, and then fed into a six-layer Clustering Transformer Encoder. Each layer comprises a clustering attention mechanism with 8 heads and a feed-forward network. Notably, the clustering attention block, proposed in \cite{zheng2020end}, computes the prototype of the queries in each cluster, reducing the number of queries from $N$ to $K$. This reduces the computational complexity of the attention map from $O(N^2)$ to $O(NK)$ compared to traditional ViT architectures \cite{dosovitskiy2020image}. For our 3D classification task, the clusters within the model correspond to the clusters assigned to each slice, simplifying queries of similar slices into single features of dimension $M$. The final result of the attention and values consists of $K$ updated vectors, which are then broadcasted back to the $N$ updated slices denoted as $s^{*}$ by replicating each feature $s_k^{*}$ into slices assigned to cluster $k$. Besides computational efficiency, the clustering mechanism also reduces the final updated features in $s^*$ by replicating the prototypes into high-dimensional data, effectively performing dimension reduction and hence mitigating overfitting issues.

Following the Clustering Transformer Encoder, the Diff3Dformer processes the updated features obtained from global learning through a linear layer to obtain the risk score for each slice denoted as $r$. After the final layer of slice fusion, the model generates a single score as the patient-level score. For our 3D classification task, the clustering ViT model is trained using cross-entropy loss.

\subsection{Interpretable Slice-Sequence Fusion}

The fusion of slice sequences plays a pivotal role in consolidating information to generate the final patient-level decision. Traditionally, this fusion is accomplished using various pooling methods or linear regression in conventional 3D analysis techniques.
In order to avoid potential overfitting resulting from dense layer and the direct averaging of patch levels, which may disregard the varying importance of individual slices in classification tasks, we propose an interpretable 3D decision-making approach, which considers the existence of different prototypes, the quantification of various clusters, and the diverse contributions of slice patterns to the final task. This can be formulated as:
\begin{equation}\label{patient_score}
R =\sum_{k=1}^{K}A_{k}\overline{r}_{k}q_{k}.
\end{equation}

Here, $A_k$ represents the global cluster attention, emphasizing the significance of the presence of a specific cluster for the final task, which remains consistent across all patients. $q_k$ indicates the ratio of the number of slices in each cluster to the total number of slices from a patient, simulating the lesion extent, while $\overline{r}_{k}$ denotes the average slice risk within cluster $k$ for each individual.

\section{Experiment}
\subsection{Dataset}
To validate the effectiveness of the proposed method on small datasets across different medical tasks, we evaluated the performance of the Diff3Dformer model in both diagnosis and prognosis tasks using two 3D datasets: COVID-19 and fibrotic lung disease (FLD). Specifically, we validated the performance of our model on the CC-CCII \cite{zhang2020clinically} dataset to tackle the classification of novel coronavirus pneumonia (NCP) and common pneumonia (CP), and on the FLD dataset for a binary prognostic task to predict the 1-year mortality of FLD patients.

\noindent \textbf{Clean-CC-CCII}:
The Clean-CC-CCII dataset is a publicly available dataset of CT volumes consisting of three different categories: NCP, CP, and normal patients, which is constructed by preprocessing and restructuring the CC-CCII dataset \cite{zhang2020clinically} in \cite{he2020benchmarking}. The Clean-CC-CCII dataset contains 3,993 scans from 2,698 patients. In this study, we perform a binary classification task of NCP and CP classes, including 1519 scans from 1047 NCP patients and 1549 scans from 824 CP patients. In our experiments, We randomly divided the scans into training data (2455 scans), validation data (306 scans), and test data (307 scans).
\renewcommand{\thefootnote}{1}

\noindent\textbf{Fibrotic lung disease}: The FLD dataset is the public dataset from OSIC\footnote{https://www.osicild.org/}, comprising 27 patients who died within one year and 704  patients who survived beyond one year during their hospitalisation. We reserve 20\% of patients for validation, and use the remaining for training. An in-house external test dataset is obtained from Australia, consisting of 501 CT scans, with 43 patients who died within one year and 458 patients who survived beyond one year. \\ 

\subsection{Implementation Details}
For representation learning, the Adam optimizer \cite{kingma2014adam} with a batch size of 64 is used to optimize the diffusion autoencoder, and the learning rate is set to 1$e^{-4}$. The input size of the diffusion autoencoder is 256 $\times$ 256. It is trained by 93967 slices generated from the 3D OSIC dataset. We trained the model using 8 V100 GPUs for 100 epochs. The number of clusters $K$ in the spherical K-means method is set to 64. The clustering ViT model is trained using the Adam optimizer \cite{kingma2014adam} on two RTX3090 GPUs with a batch size of 4 and a learning rate of $1e^{-4}$ for 100 epochs. The dimensional size $M$ is set to 512 and the dropout rate is set to 0.1. The area under the curve (AUC), accuracy, sensitivity, specificity, and F1 score were used as metrics for evaluating the performance of classification. 

\subsection{Experiment Results}
\begin{figure}[t]
\subfigure{
\centerline{\includegraphics[width=1\linewidth]{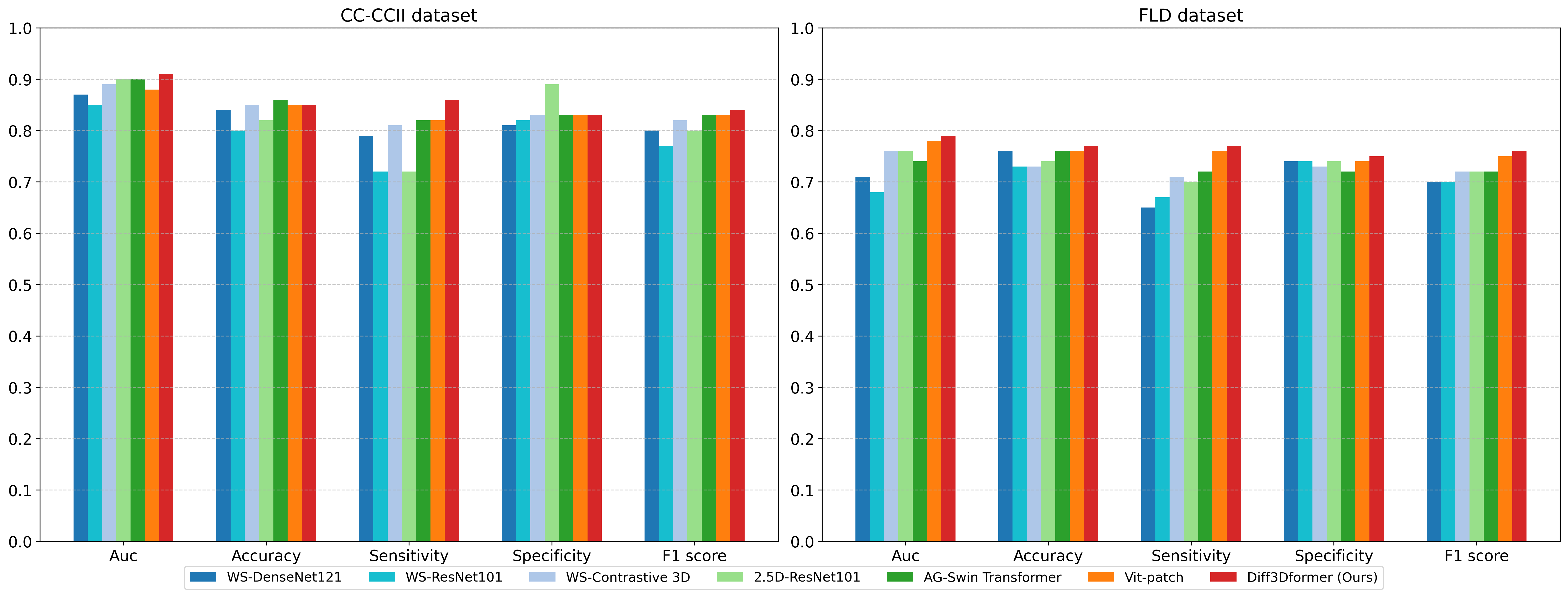}}  
}
\caption{The comparison results of different methods on CC-CCII and FLD datasets}
\label{result}
\end{figure}

In this study, we compared Diff3Dformer model with other 3D CNN-based methods and transformer-based methods which have open-source code. The comparison results on the two datasets are presented in Fig.~\ref{result}. The 3D CNN-based methods includes WS-DenseNet121 \cite{he2020benchmarking}, WS-ResNet101 \cite{he2020benchmarking},  WS-Contrastive 3D \cite{hou2021cmc}, 2.5D-ResNet101 \cite{harmon2020artificial}, and the 3D transformer-based methods includes AG-Swin Transformer \cite{zhang2021mia} and ViT-patch \cite{zhao2022prognostic}. The experiment setting of these methods can be found in Supplementary Table .\ref{tab:setting}.

\subsubsection{The proposed method outperformed other transformer-based methods.}
Compared to the AG-Swin Transformer and ViT-patch methods, our proposed model achieves superior performance in terms of AUC, sensitivity, and F1 score on the CC-CCII dataset, while demonstrating comparable performance on other metrics. On the smaller FLD dataset, the proposed method significantly enhances performance across all metrics, unlike other transformer-based methods, which exhibit sensitivity to dataset size and fail to produce satisfactory results on smaller datasets. These findings suggest that our model effectively mitigates the requirement for large datasets typically needed by transformer-based methods and demonstrates greater robustness with limited data.

\subsubsection{The proposed method consistently outperformed different type of 3D classification models.}  Comparing the F1 score on both datasets, we observe that CNN-based methods also tend to perform worse on the extremely small FLD dataset. Specifically, WS-based methods perform the worst due to the susceptibility to overfitting issues. These challenges can be partly addressed by learning a more generalizable representation through contrastive learning techniques, as demonstrated in the WS-Contrastive method and resampling methods in the 2.5D method, where the WS method outperforms the 2.5D method through a more comprehensive analysis.
In comparison to these methods on the FLD dataset, both patch-ViT and Diff3Dformer, which utilize the Clustering ViT architecture, outperform those CNN-based methods, indicating their effectiveness in reducing overfitting issues. Moreover, Diff3Dformer outperforms patch-ViT by leveraging slice-sequence analysis on both datasets.

\subsubsection{Ablation study.} 
To investigate the effectiveness of components in the proposed model on the small dataset, we conduct some experiments on both two datasets as shown in Table ~\ref{tab:ablation}. Contrastive learning  \cite{pmlr-v97-huang19b} is another useful self-supervised learning method to learn image representation, and we also remove the clustering attention in the clustering ViT model and compare their performance with the proposed model. The model without clustering attention is identical to the original ViT \cite{DBLP:journals/corr/abs-2010-11929}. The comparison between No.1 and No.2, as well as No.3 and No.4, demonstrates that the diffusion model achieves better representations compared to the contrastive learning method. When comparing No.1 to No.3 and No.2 to No.4, it is evident that clustering attention significantly improves the performance of ViT on both datasets, confirming that clustering attention effectively addresses the overfitting problem in transformer-based methods.

\begin{table*}[t]
\caption{Ablation studies on CC-CCII and FLD datasets.}
\label{tab:ablation}
\centering 
\resizebox{\textwidth}{!}{
\begin{tabular}{lll|lllllllll}
\toprule[1.5pt] 
\multicolumn{1}{c|} {}& 
\multicolumn{1}{c|} {}& \multicolumn{5}{c|}{CC-CCII} & \multicolumn{5}{c}{FLD} \\ \hline

\multicolumn{1}{c|}{No.}&  \multicolumn{1}{c|}{Ablation Setting}& \multicolumn{1}{c|}{AUC} & \multicolumn{1}{c|}{Accuracy} & \multicolumn{1}{c|}{Sensitivity} & \multicolumn{1}{c|}{Specificity} &  \multicolumn{1}{c|}{F1 Score} & \multicolumn{1}{c|}{AUC} & \multicolumn{1}{c|}{Accuracy} & \multicolumn{1}{c|}{Sensitivity} & \multicolumn{1}{c|}{Specificity} & \multicolumn{1}{c}{F1 Score} \\ \hline

\multicolumn{1}{c|}{1.}&\multicolumn{1}{c|}{Contrastive + ViT}& 
\multicolumn{1}{c|}{0.83} & \multicolumn{1}{c|}{0.82} & \multicolumn{1}{c|}{0.77} & \multicolumn{1}{c|}{0.79} & \multicolumn{1}{c|}{0.78} & \multicolumn{1}{c|}{0.75} & \multicolumn{1}{c|}{0.74} & \multicolumn{1}{c|}{0.68} & \multicolumn{1}{c|}{0.77} &  \multicolumn{1}{c}{0.72} \\ \hline

\multicolumn{1}{c|}{2.}&\multicolumn{1}{c|}{Diffusion + ViT}& 
\multicolumn{1}{c|}{0.84} & \multicolumn{1}{c|}{0.84} & \multicolumn{1}{c|}{0.78} & \multicolumn{1}{c|}{0.81} & \multicolumn{1}{c|}{0.79} & \multicolumn{1}{c|}{0.76} & \multicolumn{1}{c|}{0.76} & \multicolumn{1}{c|}{0.68} & \multicolumn{1}{c|}{0.79} &  \multicolumn{1}{c}{0.73}\\ \hline

\multicolumn{1}{c|}{3.}&\multicolumn{1}{c|}{Contrastive + clustering ViT}& 
\multicolumn{1}{c|}{0.88} & \multicolumn{1}{c|}{0.84} & \multicolumn{1}{c|}{0.81} & \multicolumn{1}{c|}{0.83} & \multicolumn{1}{c|}{0.82}& \multicolumn{1}{c|}{0.78} & \multicolumn{1}{c|}{0.75} & \multicolumn{1}{c|}{0.75} & \multicolumn{1}{c|}{0.74} &  \multicolumn{1}{c}{0.74} \\ \hline

\multicolumn{1}{c|}{4.}&\multicolumn{1}{c|}{Diffusion + clustering ViT}& 
\multicolumn{1}{c|}{0.91} & \multicolumn{1}{c|}{0.85} & \multicolumn{1}{c|}{0.86} & \multicolumn{1}{c|}{0.83} & \multicolumn{1}{c|}{0.84}& \multicolumn{1}{c|}{0.79} & \multicolumn{1}{c|}{0.77} & \multicolumn{1}{c|}{0.77} & \multicolumn{1}{c|}{0.75} &  \multicolumn{1}{c}{0.76}\\
\bottomrule[1.5pt]
\end{tabular}
}
\end{table*}

\subsubsection{Interpretable results.} 
Based on Eqn. \eqref{patient_score}, we can identify the most influential cluster contributing to the final score $R$ for each individual by vectorizing the feature $A_{k}\overline{r}_{k}$ for each cluster. The heatmap in Supplementary Fig. \ref{fig:new_heatmap} represents the contribution of the cluster to the final patient-level risk score $R$ on the FLD dataset, where the panels from left to right depict the $A_
{k}\overline{r}_{k}$ vectors for patients arranged in decreasing order of $R$ value. The rationale behind each patient's final prediction: the red cube highlights clusters contributing to high-risk scores, while blue indicates a lower risk.  From this visualization, we can see that patients with different prediction results are highly disentangled, and the contributing patterns are clearly delineated for each patient. The most influential clusters across the dataset are determined by comparing the average $A_{k}\overline{r}_{k}$ values between the two classes with different predictions. The ranking of clusters by contribution to the `mortality in one year' class on the FLD dataset is shown in  Supplementary Fig. \ref{fig:new_rank} and the most representative slice patterns are provided in Supplementary Fig. \ref{fig:example_figure}, which show that the model can identify common clusters within each class group, enabling us to pinpoint most significant features by visualizing the most frequently contributing clusters among patients.

\section{Conclusion}
In this paper, we propose Diff3Dformer specifically tailored to overcome the challenges encountered in classifying 3D CT scans using small medical image datasets, outperforming both CNN-based and Transformer-based methods. Leveraging Diffusion-based slice-sequence representations empowers Transformer architecture for high-dimensional 3D volume data, and enhances classification accuracy with its rich and meaningful feature representation. Experimental results demonstrate the superior performance of our proposed method across various scales of small datasets and medical image classification tasks.

\subsubsection{\ackname} This study was supported in part by the ERC IMI (101005122), the H2020 (952172), the MRC (MC/PC/21013), the Royal Society (IEC/NSFC/211235), the NVIDIA Academic Hardware Grant Program, the SABER project supported by Boehringer Ingelheim Ltd, Wellcome Leap Dynamic Resilience, NIHR Imperial Biomedical Research Centre, and the UKRI Future Leaders Fellowship(MR/V023799/1).

\subsubsection{\discintname}
The authors have no competing interests to declare that are relevant to the content of this article.

\bibliographystyle{splncs04}
\bibliography{reference}

\newpage

\section{Supplementary materials}

\begin{table}
  \centering
  \captionsetup{labelfont=bf}
  \caption{The experiment setting of the methods for comparison in the paper. $z$ is the number of slices and $p$ is the number of patches cropped from the whole CT scan.}
  \resizebox{0.8\linewidth}{!}{
  \begin{tabular}{cccccc}
    \toprule
   \textbf{Model Name} & \textbf{Learning Rate} & \textbf{Batch Size} & \textbf{Optimizer} & \textbf{Hardware} & \textbf{Input Size} \\
    \midrule
    WS-DenseNet121 & 1e-3 & 32 & Adam Optimizer & One RTX3090 & 64 × 128 × 128 \\
    WS-ResNet101 & 1e-3 & 32 & Adam Optimizer & One RTX3090 & 64 × 128 × 128 \\
    WS-Contrastive 3d & 1e-4 & 4 & Adam Optimizer & Two RTX3090 & 64 × 256 × 256 \\
    2.5D-ResNet101 & 1e-4 & 8 & Adam Optimizer & Two RTX3090 & 8 × 256 × 256 \\
    AG-Swin Transformer & 1e-4 & 2 & Adam Optimizer & Two RTX3090 &  $z$ × 224 × 224 \\
    ViT-patch & 1e-5 & 4 & Adam Optimizer & Two RTX3090 & $p$ × 64 × 64 \\
    \bottomrule
  \end{tabular}
  }
  \label{tab:setting}
\end{table}

\begin{figure}
    \centering
   \includegraphics[width=\textwidth]{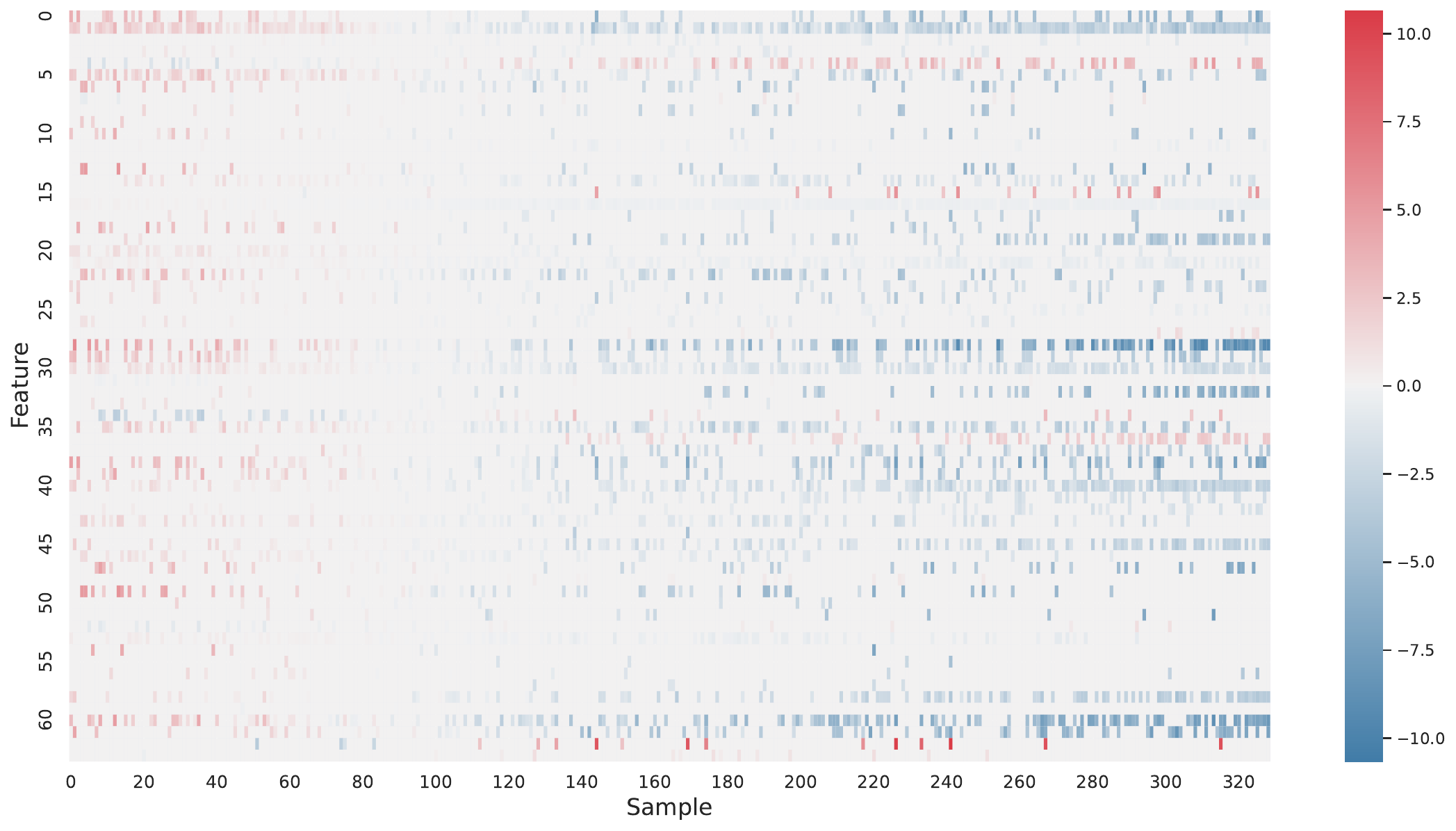}
     \captionsetup{labelfont=bf}
    \caption{The heatmap represents the contribution of the cluster to the final patient-level risk 
score $R$ on the FLD dataset. Patients ranked from highest to lowest risk score $R$ on the 
horizontal axis from the left to right and 64 clusters on the vertical axis.}
    \label{fig:new_heatmap}
\end{figure}

\begin{figure}
    \centering
    \includegraphics[width=0.8\textwidth]{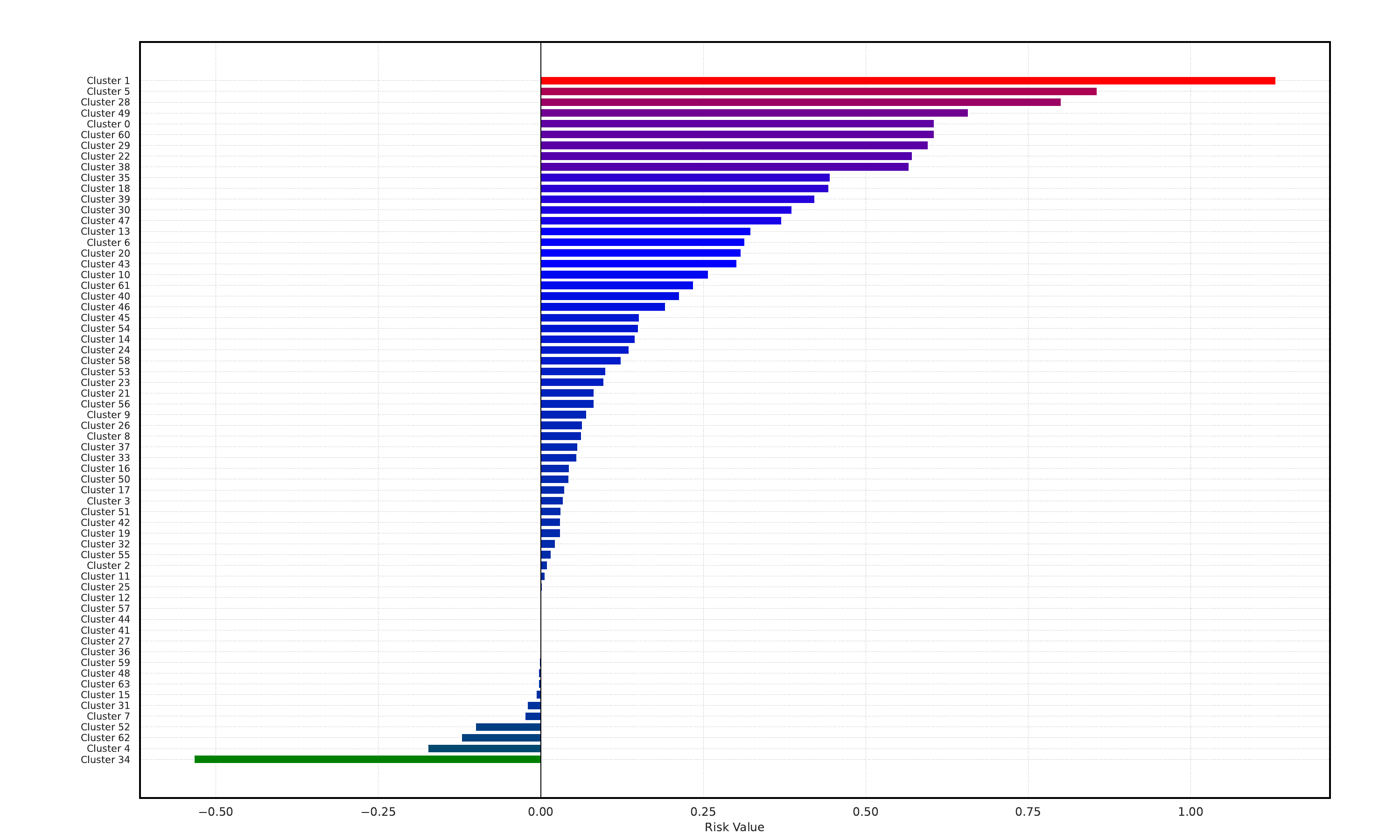}
      \captionsetup{labelfont=bf}
    \caption{ Cluster ranking by contribution to the `mortality in one year' class on the FLD
dataset.}
    \label{fig:new_rank}
\end{figure}

\vspace{-10mm}
\begin{figure*}[htbp]
    \centering
    \begin{minipage}{.05\textwidth}
        \rotatebox{90}{\scriptsize{\textbf{Cluster 1}}}
    \end{minipage}%
    \begin{minipage}{0.8\textwidth}
        \includegraphics[width=0.185\linewidth]{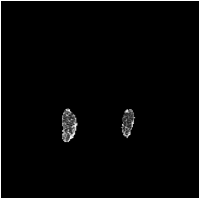}%
        \hfill
        \includegraphics[width=0.185\linewidth]{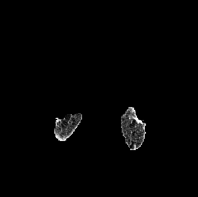}%
        \hfill
        \includegraphics[width=0.185\linewidth]{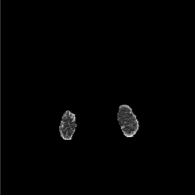}%
        \hfill
        \includegraphics[width=0.185\linewidth]{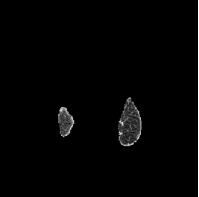}%
        \hfill
        \includegraphics[width=0.185\linewidth]{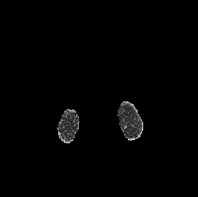}%
    \end{minipage}
    \vspace{3mm}
    
        \begin{minipage}{.05\textwidth}
        \rotatebox{90}{\scriptsize{\textbf{Cluster 5}}}
    \end{minipage}%
    \begin{minipage}{0.8\textwidth}
        \includegraphics[width=0.185\linewidth]{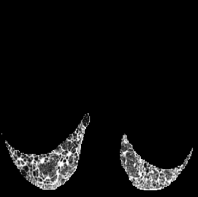}%
        \hfill
        \includegraphics[width=0.185\linewidth]{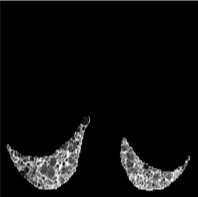}%
        \hfill
        \includegraphics[width=0.185\linewidth]{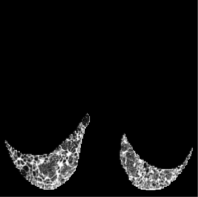}%
        \hfill
        \includegraphics[width=0.185\linewidth]{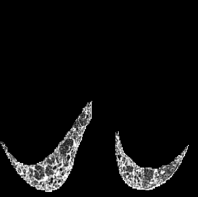}%
        \hfill
        \includegraphics[width=0.185\linewidth]{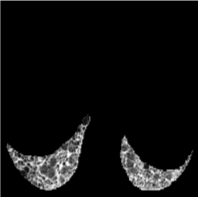}%
    \end{minipage}
        \vspace{3mm}
    
        \begin{minipage}{.05\textwidth}
        \rotatebox{90}{\scriptsize{\textbf{Cluster 28}}}
    \end{minipage}%
    \begin{minipage}{0.8\textwidth}
        \includegraphics[width=0.185\linewidth]{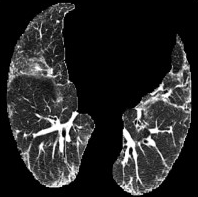}%
        \hfill
        \includegraphics[width=0.185\linewidth]{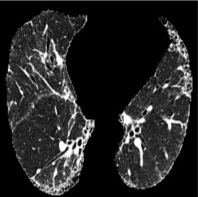}%
        \hfill
        \includegraphics[width=0.185\linewidth]{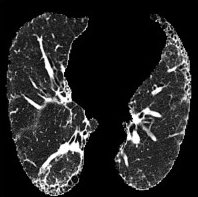}%
        \hfill
        \includegraphics[width=0.185\linewidth]{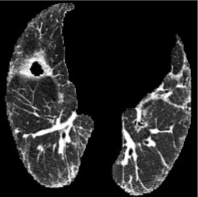}%
        \hfill
        \includegraphics[width=0.185\linewidth]{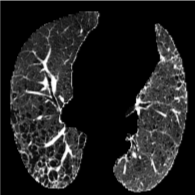}%
    \end{minipage}
            \vspace{3mm}
    
        \begin{minipage}{.05\textwidth}
        \rotatebox{90}{\scriptsize{\textbf{Cluster 49}}}
    \end{minipage}%
    \begin{minipage}{0.8\textwidth}
        \includegraphics[width=0.185\linewidth]{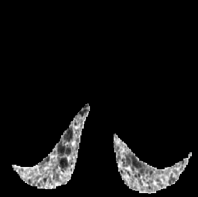}%
        \hfill
        \includegraphics[width=0.185\linewidth]{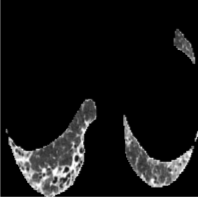}%
        \hfill
        \includegraphics[width=0.185\linewidth]{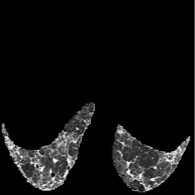}%
        \hfill
        \includegraphics[width=0.185\linewidth]{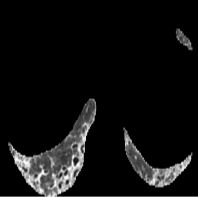}%
        \hfill
        \includegraphics[width=0.185\linewidth]{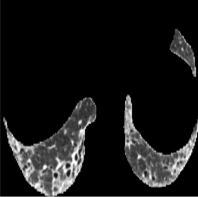}%
    \end{minipage}
    
    \vspace{3mm} 
    
        \begin{minipage}{.05\textwidth}
        \rotatebox{90}{\scriptsize{\textbf{Cluster 0}}}
    \end{minipage}%
    \begin{minipage}{0.8\textwidth}
        \includegraphics[width=0.185\linewidth]{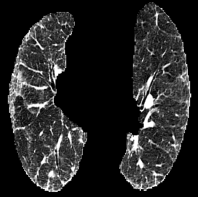}%
        \hfill
        \includegraphics[width=0.185\linewidth]{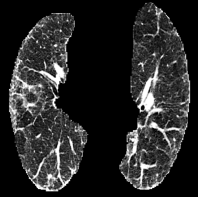}%
        \hfill
        \includegraphics[width=0.185\linewidth]{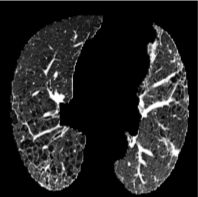}%
        \hfill
        \includegraphics[width=0.185\linewidth]{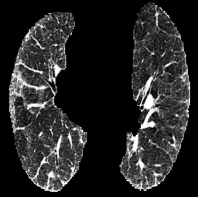}%
        \hfill
        \includegraphics[width=0.185\linewidth]{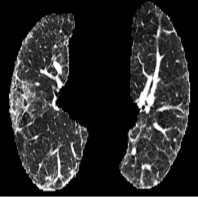}%
    \end{minipage}
    
  \captionsetup{labelfont=bf}
    \caption{Visualization of the representative slices of high-risk clusters on the FLD dataset. The representative slices are those closest to the centroids of the cluster.}
    \label{fig:example_figure}
\end{figure*}

\end{document}